\documentstyle[12pt]{article}

\normalbaselines
\begin{document}
\begin{center}
\vspace*{1.0cm}

{\LARGE{\bf
On Interbasis Expansion for Isotropic Oscillator
on Two-Dimensional Sphere
}}

\vskip 1.5cm

{\large {\bf Ye.M. Hakobyan, G.S. Pogosyan}}

\vskip 0.5 cm

{\sl Laboratory of Theoretical Physics, \\Joint Institute for
Nuclear Research,}\\
{\sl Dubna, Moscow Region 141980, Russia}

\end{center}

\vspace{1 cm}

\begin{abstract}
In this article we analyze the isotropic oscillator system
on the two-dimensional sphere in the spherical systems of
coordinates. The expansion coefficients for transitions
between three spherical bases of the oscillator are
calculated. It is shown that these coefficients are expressed
through the Clebsch-Gordan coefficients for SU(2) group
analytically continued to real values of their argument.

\end{abstract}

\vspace{1 cm}

\section{Introduction}

The present article is devoted to the oscillator system
on the two-dimensional sphere $s_1^2+s_2^2+s_3^2=R^2$,
which is also known  as a Higgs oscillator \cite{HIG}
\begin{eqnarray}
\label{I1}
V = \frac{\alpha^2 R^2}{2}\, \frac{s_1^2+s_2^2}{s_3^2},
\end{eqnarray}
where $s_i$ are the Cartesian coordinates in the ambient Euclidean
space and $R$ is a radius of sphere. As a "flat" space partner
\cite{POG}, this is a superintegrable system and has the same
properties as accidental degeneracy of the energy spectrum
\cite{HIG}, separation of variables in more than one coordinate
systems \cite{GRO1,KMP} and nontrivial realization of hidden
symmetries \cite{DASK} (see also \cite{ZHED}).

The aim of this paper is to describe of solutions to the
Schr\"odinger equation for the potential (1) in three spherical
systems of coordinates and to calculate the coefficients
of interbasis expansion between the corresponding wave functions.

\section{Quantum motion on the two dimensional sphere}

The Schr\"odinger equation on the two-dimensional sphere
has the following form:
\begin{eqnarray}
{H} \Psi = \left[- \frac{1}{2} \Delta_{LB} + V\right]
\Psi = E \Psi
\end{eqnarray}
where $\Delta_{LB}$ is the Laplace--Beltrami operator
\begin{equation}
\Delta_{LB} = \frac{1}{R^2} (L^2_1 + L^2_2 + L^2_3).
\end{equation}
and $L_i$ are the generators of the Lie algebra $o(3)$
\begin{eqnarray}
L_i = -\epsilon_{ikj}s_k\frac{\partial}{\partial s_j},\quad
[L_i, L_k]=\epsilon_{ikj} L_j,\quad i,k,j = 1,2,3
\end{eqnarray}
For $V=0$ the separated eigenfunctions of the Laplace-Beltrami
operator satisfy
\begin{eqnarray}
\label{SEP1}
\Delta_{LB} \Psi = - \frac{l(l+1)}{R^2}, \quad
I \Psi = k \Psi,   \quad
\Psi_{lk}(\alpha, \beta) = \psi_{lk}(\alpha) \psi_{lk}(\beta)
\end{eqnarray}
where $I$ is a second order operator in the enveloping algebra of
o(3)
\begin{eqnarray}
I = a_{ik} L_i L_k, \qquad a_{ik} = a_{ki}
\end{eqnarray}
The matrix $a_{ik}$ can be diagonalized to give \cite{WIN}
\begin{eqnarray}
I(a_1, a_2, a_3) = a_1 L_1^2 + a_2 L_2^2 + a_3 L_3^2
\end{eqnarray}
When all three eigenvalues $a_i$ are different, the separable
coordinates in (\ref{SEP1}) are elliptic \cite{PW}.
If the two eigenvalues of $a_i$ are equal, e.g. $a_1=a_2\not=a_3$
or $a_1\not=a_2=a_3$, or $a_1=a_3\not=a_2$ we can transform
the operator $I$ into the operators:
$I(0,0,1) = L_3^2$, $I(0,1,0)=L_2^2$, or $I(1,0,0) = L_1^2$.
Thus, the corresponding separable coordinates on $S_2$ are the
three type of spherical coordinates
\begin{eqnarray}
\label{COOR1}
\begin{array}{llll}
s_1 &=\, R\sin\theta\cos\varphi
    &=\, R\cos\theta'
    &=\, R\sin\theta''\sin\varphi'', \\
s_2 &=\, R\sin\theta\sin\varphi
    &=\, R\sin\theta'\cos\varphi'
    &=\, R\cos\theta'',              \\
s_3 &=\, R\cos\theta
    &=\, R\sin\theta'\sin\varphi'
    &=\, R\sin\theta''\cos\varphi''
\end{array}
\end{eqnarray}
where $\varphi \in [0, 2\pi),$ $\theta \in (0, \pi)$. The
eigenfunctions of the three sets of operators $ \{\Delta_{LB}, L_i\}$
are the usual spherical functions on $S_2$:
\begin{eqnarray}
\Delta_{LB}\, Y_{lm_i}(\theta, \varphi)
= - \frac{l(l+1)}{R^2}\,  Y_{lm_i}(\theta, \varphi)
\quad
L_i^2 Y_{lm_i}(\theta, \varphi)  =  m_i^2 Y_{lm_i}(\theta, \varphi)
\end{eqnarray}
Geometrically, the spherical coordinates (\ref{COOR1}) are
connected with each other by rotation which may be expressed through
the Euler angles $(\alpha,\beta,\gamma)$ in accordance with the
relations \cite{V}
\begin{eqnarray*}
\cos\theta'
&=& \cos\theta \cos\beta + \sin\theta\sin\beta\cos(\varphi-\alpha)
\\[2mm]
\cot(\varphi'+ \gamma)
&=& \cot(\varphi-\alpha)\cos\beta
-\frac{\cot\theta\sin\beta}{\sin(\varphi-\alpha)}
\end{eqnarray*}
Correspondingly, the spherical functions $Y_{lm}(\theta,\varphi)$
are transformed by the formulae \cite{V}
\begin{eqnarray*}
Y_{l,m'}(\theta',\varphi')&=&\sum_{m=-l}^{l}D_{mm'}^l
(0,\frac{\pi}{2},\frac{\pi}{2})
Y_{l,m}(\theta,\varphi),
\\[2mm]
Y_{l,m''}(\theta'',\varphi'')&=&\sum_{m=-l}^{l}D_{mm''}^l
(\frac{\pi}{2},\frac{\pi}{2},0)
Y_{l,m}(\theta,\varphi),
\\[2mm]
Y_{l,m''}(\theta'',\varphi'')&=&\sum_{m'=-l}^{l}D_{m'm''}^l
(0,\frac{\pi}{2},\frac{\pi}{2})
Y_{l,m'}(\theta',\varphi'),
\end{eqnarray*}
where $D_{m_1, m_2}^l (\alpha,\beta,\gamma)$ - are
the Wigner $D$-functions.

\section{Higgs oscillator on the two-dimensional sphere}

\subsection{Solution to the Schr\"odinger equation}

{\it 3.1} \, The oscillator potential (1) in
the spherical coordinate $(\theta,\varphi)$ is
\begin{eqnarray}
\label{P1}
V = \frac{\alpha^2 R^2}{2}\, \frac{s_1^2+s_2^2}{s_3^2}
= \frac{\alpha^2 R^2}{2}\tan^2\theta
\end{eqnarray}
and the Schr\"odinger equation (2) has the following form:
\begin{eqnarray}
\label{SCH1}
\frac{1}{\sin\theta}\frac{\partial}{\partial \theta}
\sin\theta\frac{\partial\Psi}{\partial \theta}
+\frac{1}{\sin^2\theta}\frac{\partial^2\Psi}{\partial\varphi^2}
+2 R^2\left[E-
\frac{\alpha^2 R^2}{2}\tan^2\theta
\right]\Psi = 0
\end{eqnarray}
Choosing the wave function according to
\begin{eqnarray}
\label{WF1}
\Psi(\theta, \varphi) =
\frac{Z(\theta)}{\sqrt{\sin\theta}}
\,
\frac{e^{i m \varphi}}{\sqrt{2\pi}},
\qquad
m \in {\mbox{\bf Z}},
\end{eqnarray}
after the separation of variables in
equation (\ref{SCH1}) we come to the P\"oschl-Teller - type
equation:
\begin{eqnarray}
\label{PT}
\frac{d^2 Z}{d\theta^2}+\left[\varepsilon
-\frac{m^2-\frac14}{\sin^2\theta}
-\frac{\nu^2-\frac14}{\cos^2\theta}
\right] Z = 0
\end{eqnarray}
where $\nu =\sqrt{\alpha^2 R^4+\frac{1}{4}}$ and
$\varepsilon = 2 R^2E+\alpha^2 R^4+\frac14$.
The solution of the above equation orthonormalised
in the interval $\theta\in [0,\pi/2]$ is
\begin{eqnarray}
\label{}
Z(\theta) \equiv Z_{n_rm}
(\theta)
&=&
\sqrt{\frac{2(2n_r+|m|+\nu+1)(n_r)!\Gamma(n_r+|m|+\nu+1)}
{(n_r+|m|)!\Gamma(n_r+\nu+1)}}
\nonumber
\\[2mm]
&\cdot&
(\sin\theta)^{|m|+\frac12}
(\cos\theta)^{\nu+\frac{1}{2}}
P_{n_r}^{(|m|, \nu)}(\cos 2\theta)
\end{eqnarray}
where $P_{n}^{(\alpha, \beta)}(x)$ are the Jacobi polynomials \cite{BE}
and the energy $E$ takes the values
\begin{eqnarray}
\label{EN}
E_n=\frac{1}{2R^2}\left[(n+1)(n+2) + (2\nu-1)(n+1)\right]
\end{eqnarray}
where $n_r$ is a "radial" quantum number and $n = 2n_r+|m|$
is the principal quantum number. The degree of degeneracy
of the energy spectrum, like the flat two-dimensional
oscillator system, is equal to $2n+1$. Note also that in
contraction limit when $R\rightarrow \infty$, we have
$\nu\sim\alpha R^2$ and from formula (\ref{EN})
the energy spectrum for two-dimensional circular oscillator
is restored \cite{FLUG}.

\vspace{0.3cm}
\noindent
{\it 3.2} \, In the second spherical coordinate
$(\theta', \varphi')$ the potential (1) has the form
\begin{eqnarray}
\label{P2}
V = \frac{\alpha^2 R^2}{2}\left(\frac{1}
{\sin^2\theta' \sin^2\varphi'}-1\right)
\end{eqnarray}
After the substitution
\begin{eqnarray}
\label{WF2}
\Psi(\theta', \varphi') = \frac{1}{\sqrt{\sin\theta'}}
S(\theta')S(\varphi')
\end{eqnarray}
we come to the system of differential equations
\begin{eqnarray}
\label{SCH2}
\frac{d^2 S}{d\theta^{'2}}+ \left[\varepsilon
-\frac{A^2-\frac{1}{4}}{\sin^2\theta'}\right] S = 0
\qquad
\frac{d^2 S}{d\varphi^{'2}} + \left[A^2
-  \frac{\nu^2-\frac{1}{4}}{\sin^2\varphi'}\right]S = 0
\end{eqnarray}
where $A$ is the separation constant. Solving  equations
(\ref{SCH2}) we obtain
\begin{eqnarray}
\label{}
A = n_1+\nu+\frac{1}{2},
\quad
\varepsilon = \left(n_2+A+\frac{1}{2}\right)^2 =
(n_1+n_2+\nu+1)^2 = (n+\nu+1)^2
\end{eqnarray}
where $n_1, n_2 \in {\bf N}$ and the principal quantum number
$n=n_1+n_2$, so that the energy spectrum is given by equation
(\ref{EN}). The orthonormalized eigenfunctions
$\Psi(\theta', \varphi')$ could be written as
\begin{eqnarray}
\label{SOL2}
\Psi_{n_1 n_2}(\theta', \varphi') = \frac{1}{\sqrt{\sin\theta'}}
S_{n_2}^A (\theta')
S_{n_1}^\nu(\varphi')
\end{eqnarray}
where
\begin{eqnarray}
\label{SOL1}
S_n^a(\varphi)
&=&
\frac{\Gamma(a+1)\Gamma(n+a+\frac12)}{\Gamma(n+a+1)}
\sqrt{\frac{(n+a+1/2)n!}{\pi \Gamma(n+2a+1)}} \,
(\sin\varphi )^{{1\over 2}+a}
C^{a+\frac12}_{n}(\cos\varphi)
\end{eqnarray}
and $C_n^{\lambda}$ are the Gegenbauer polynomials \cite{BE}.
Finally, note that the operator characterizing the
separation solutions in this coordinate system is
\begin{eqnarray}
\label{}
J_1\Psi_{n_1 n_2} = \left(
\frac{\partial^2}{\partial \varphi'^2}
-\frac{\nu^2-\frac{1}{4}}{\sin^2\varphi'}\right)
\Psi_{n_1 n_2} =
\left[L_1^2- (s_2^2+s_3^2)
\frac{\nu^2-\frac{1}{4}}{s_3^2}\right]
\Psi_{n_1 n_2} =
- A^2 \Psi_{n_1 n_2}
\end{eqnarray}

\vspace{0.3cm}
\noindent
{\it 3.3} \, For the potential (1) in the coordinate system
$(\theta'', \varphi'')$ we have
\begin{eqnarray}
\label{P3}
V = \frac{\alpha^2 R^2}{2}\left(\frac{1}
{\sin^2\theta'' \cos^2\varphi''}-1\right)
\end{eqnarray}
The orthonormalized solution to the Schr\"odinger equation (2)
have the following form:
\begin{eqnarray}
\label{WF3}
\Psi_{l_1 l_2}(\theta'', \varphi'') =
\frac{1}{\sqrt{\sin\theta''}}
S_{l_1}^\nu (\varphi''+ \frac{\pi}{2})
\,
S_{l_2}^B (\theta'')
\end{eqnarray}
where $l_1, l_2 \in {\bf N}$, the principal quantum number
$n=l_1+l_2$ and the constant $B=l_1+\nu+\frac12$. For the
energy spectrum we come to expression (\ref{EN}) and
the wave function $S_n^a(\theta)$ is given by formula (\ref{SOL1}).

The additional operator describing this solution
and separation is
\begin{eqnarray}
\label{}
J_2\Psi_{l_1 l_2}=
\left(\frac{\partial^2}{\partial \varphi''^2}
- \frac{\nu^2-\frac14}{\cos^2\varphi''}\right)
\Psi_{l_1 l_2}=
\left[L_2^2- (s_1^2+s_3^2)\frac{\nu^2-\frac14}{s_3^2}
\right] \Psi_{l_1 l_2}
= - B^2 \Psi_{l_1 l_2}
\end{eqnarray}

\subsection{Algebra}

If we take the constant of motion in the form
\begin{eqnarray*}
{\tilde J_3} = L_3, \qquad
{\tilde J_1} = L_1^2 -  \alpha^2 R^4 \, \frac{s_2^2}{s_3^2},
\quad
{\tilde J_2} = L_2^2 -  \alpha^2 R^4 \, \frac{s_1^2}{s_3^2},
\end{eqnarray*}
we have the Hamiltonian
\begin{eqnarray*}
H = - \frac{1}{2 R^2}\bigg[{\tilde J_1}+ {\tilde J_2}+
{\tilde J_3^2}\bigg],
\end{eqnarray*}
and the commutator relations
\begin{eqnarray}
[\tilde J_1, \tilde J_2] &=&
\{L_1,\{L_2,L_3\}\}+2 \alpha^2 R^4\left(
\frac{s_2^2-s_1^2}{s_3^2}+\frac{2 s_1 s_2}{s_3^2}L_3
\right)
\nonumber
\\[2mm]
[\tilde J_1, \tilde J_3] &=&
-\{L_1,L_2\}-2 \alpha^2 R^4
\frac{s_1s_2}{s_3^2}
\nonumber
\\[2mm]
[\tilde J_2, \tilde J_3] &=&
\{L_1,L_2\}+2 \alpha^2 R^4
\frac{s_1s_2}{s_3^2}
\end{eqnarray}
where $\{,\}$ is the anticommutator.
To close this algebra, we use the redefined operators
\begin{eqnarray}
S_1 = {\tilde J_3}, \qquad
S_2 = \tilde J_1-\tilde J_2, \qquad
S_3 = [S_1,S_2]
\end{eqnarray}
and derive the following relations:
\begin{eqnarray}
S_3&=&2\{L_1 , L_2\}
+  4\alpha^2 R^4 \frac{s_1s_2}{s_3^2},
\\[2mm]
[S_3, S_1]&=& 4 S_2, \quad
[S_3, S_2] = \frac{4 H S_1}{R^2}
+ 8 S_1^3
+4\left(4\alpha^2 R^4-1\right) S_1.
\end{eqnarray}
Thus, the operators $S_1, S_2, S_3$ a generate nonlinear algebra, the
so-called cubic or Higgs algebra.

\section{Interbasis expansions}

Let us now consider interbasis expansion between two spherical wave
functions
\begin{eqnarray}
\label{EXP1}
\Psi_{n_1,n_2}(\theta',\varphi')=
\sum_{m = - n}^{n}
W_{n_1 n_2}^{m}
\Psi_{n, m}(\theta,\varphi)
\end{eqnarray}
To calculate an explicit form of the expansion coefficients
$W_{n_1 n_2}^{m}$ it is sufficient to use the orthogonality
for the wave function on one of the variables in the right-hand
side of (\ref{EXP1}) and to fix, at the most appropriate point,
the second variable that does not participate in integration.
Rewrite the left-hand side of (\ref{EXP1}) in the spherical
coordinates $(\theta, \varphi)$ according to the formulae
\begin{eqnarray*}
\cos\theta' = \sin\theta\cos\varphi, \quad
\cos\varphi'= \frac{\sin\theta\sin\varphi}
{\sqrt{1-\sin^2\theta\cos^2\varphi}}.
\end{eqnarray*}
Then, by substituting $\theta = \frac{\pi}{2}$ and taking into
account that
\begin{eqnarray*}
C_{n}^{\lambda}(1) = \frac{\Gamma(2\lambda+n)}
{n!\Gamma(2\lambda)}
\end{eqnarray*}
we obtain an equation depending only on the variable $\varphi$.
Thus, using the orthogonality relation for the function
$e^{im\varphi}$ upon the quantum number $m$, we arrive at the
following integral representation for the coefficients
$W_{n_1,n_2}^{m}$:
$$
W_{n_1 n_2}^{m}(\nu)
= \frac{(-1)^{\frac{n-|m|}{2}}}{2^{\nu+1}\pi}
\,
\sqrt{\frac{(n_2)!\left(n_1+\nu+\frac12\right)\Gamma(n_1+2\nu+1)
(\frac{n+m}{2})!(\frac{n-m}{2})!}
{(n_1)!\Gamma(n+n_1+2\nu+2)
\Gamma(\frac{n-m}{2}+\nu+1)\Gamma(\frac{n+m}{2}+\nu+1)}}
$$
\begin{eqnarray}
\label{EXP11}
\cdot
\,
\frac{\Gamma(n_1+\nu+\frac32)\Gamma(n+\nu+1)}
{\Gamma(n+\nu+\frac32)}
\,\,
I_{n_1 n_2 m}^{\nu}
\end{eqnarray}
where
\begin{eqnarray}
\label{INT1}
I_{n_1 n_2 m}^{\nu} =
\frac{1}{\sqrt{2\pi}}
{\int_{0}^{2\pi}
(\sin\varphi)^{n_1}
C_{n_2}^{n_1+\nu+1}(\cos \varphi)
e^{-i m \varphi}
d\varphi}.
\end{eqnarray}
To calculate the integral $I_{n_1 n_2 m}^{\nu}$ it is sufficient
to write the Gegenbauer polynomial $C_{n_2}^{n_1+\nu}(\cos \varphi)$
and $(\sin\varphi)^{n_1}$ as a series in terms of the exponents.
After integration we obtain
\begin{eqnarray*}
{\int_{0}^{2\pi}
(\sin\varphi)^{k}
C_{n}^{\lambda}(\cos \varphi)
e^{-i m \varphi}
d\varphi}&=&
\frac{(-1)^{\frac{n-m}{2}} 2^{\lambda-k+\frac12}
\pi\Gamma(\lambda+n +\frac12) k!}
{n!\Gamma(\lambda+\frac12)\Gamma\left(\frac{n+k-m}{2}+1\right)
\Gamma\left(\frac{k-n+m}{2}+1\right)}
\nonumber
\\[3mm]
&\cdot&{_3F_2}\left\{\left.\matrix{
-n,\quad\,\,\,-\frac{n+k-m}{2},\quad\,\,\lambda\cr
\quad\,\,\,\,\,\,\,\,\cr
-\lambda-n+1,\,\frac{k-n+m}{2}+1\cr}\right| 1\right\}
\end{eqnarray*}
The introduction of (\ref{INT1}) into (\ref{EXP11}) gives
us the interbasis coefficients in the closed form
\begin{eqnarray}
\label{COEF1}
W_{n_1 n_2}^{m}(\nu)
&=& (-1)^{\frac{|m|-m-n_1}{2}}
\sqrt{\frac{2 (n_1+\nu+\frac{1}{2}) (n_1)!
\Gamma(n_1+2\nu+1)}
{(n_2)!\Gamma(n+n_1+2\nu+2)\Gamma(\frac{n-m}{2}+\nu+1)
\Gamma(\frac{n+m}{2}+\nu+1)}}
\nonumber
\\[3mm]
&\cdot&
\sqrt{\frac{(\frac{n+m}{2})!}{(\frac{n-m}{2})!}}
\frac{\Gamma(n+\nu+1)}{\Gamma(\frac{n_1-n_2+m}{2}+1)}\,\,\,
{_3F_2}\left\{\left.\matrix{
-n_2,\,\,\,-\frac{n-m}{2},\,\,\,n_1+\nu+1\cr
\quad\,\,\,\,\,\,\,\,\cr
-n-\nu,\,\frac{n_1-n_2+m}{2}+1\cr}\right| 1\right\}
\end{eqnarray}

The interbasis coefficients $W_{n_1 n_2}^{m}(\nu)$ could also be
expressed in term of the Clebsch--Gordon coefficients for $SU(2)$ group,
analytically continued to the real values of their arguments.
Using the formula for the Clebsch-Gordon coefficients
$C_{a, \alpha; b, \beta}^{c, \gamma}$ \cite{V}
\begin{eqnarray}
\label{CG1}
C_{a, \alpha; b, \beta}^{c,\gamma}=
      \delta_{\gamma,\alpha+\beta}
\sqrt{\frac{(a+\alpha)!(b-\beta)!(c+\gamma)!(c-\gamma)!(2c+1)}
{(a+b-c)!(a+b+c+1)!(a-\alpha)!(b+\beta)!}}
\nonumber
\\[3mm]
\frac{\sqrt{(a-b+c)!(c-a+b)!}}{(-b+c+\alpha)!(-a+c-\beta)!}\,\,\,
{_3F_2}\left\{\left.\matrix{
-a-b+c,\,\,\,-a+\alpha,\,\,\,-b-\beta\cr
-a+c-\beta+1,\,-b+c+\alpha+1\cr}\right| 1\right\}
\end{eqnarray}
and following the property of the polynomial hypergeometric function
$_3F_2$
\begin{eqnarray}
{_3F_2}\left\{\left.\matrix{
a,\,\,\,b,\,\,\,c\cr
d,\quad e\cr}\right| 1\right\}=\frac{\Gamma(d) \Gamma(d-a-b)}
{\Gamma(d-a)\Gamma(d-b)}\,\,\,
{_3F_2}\left\{\left.\matrix{
a,\,\,\,b,\,\,\,e-c\cr
a+b-d+1,\,e\cr}\right| 1\right\}
\end{eqnarray}
We can rewrite the formula (\ref{CG1}) in the form
\begin{eqnarray}
\label{CG2}
C_{a,\alpha; b, \beta}^{c, \gamma}=\delta_{\gamma,\alpha+\beta}
\sqrt{\frac{(2c+1)(b+c-a)!(b-\beta)!(c+\gamma)!(c-\gamma)!      }
{(a+b-c)!(a-b+c)!(a+b+c+1)!(a+\alpha)!(a-\alpha)!(b+\beta)!}}
\nonumber
\\[2mm]
\frac{(2a)!(c-b+\alpha)!}
{(c-b+\alpha)!(c-a-\beta)!}\,\,\,
{_3F_2}\left\{\left.\matrix{
-a-b+c,\,\,\,-a+\alpha,\,\,\,b-a+c+1\cr
-2a,\,c-a-\beta+1\cr}\right| 1\right\}
\end{eqnarray}
By comparing equations (\ref{CG2}) and (\ref{COEF1}) we finally
obtain
\begin{eqnarray}
\label{COEF2}
W_{n_1 n_2}^m(\nu)
=
(-1)^{\frac{|m|-m-n_1}{2}}
\,
C_{\frac{n+\nu}{2}, \, \frac{\nu+m}{2}; \,
\frac{n+\nu}{2}, \, \frac{\nu-m}{2}}^{n_1+\nu, \, \nu}
\end{eqnarray}
The inverse expansion of (\ref{EXP1}), namely
\begin{eqnarray}
\label{EXP2}
\Psi_{n m}(\theta, \varphi) =
\sum_{n_1 = 0}^{n}
{\tilde W_{n m}^{n_1}(\nu)}
\Psi_{n_1 n_2}(\theta', \varphi')
\end{eqnarray}
immediately follows from the orthogonality property of the
$SU(2)$ Clebsch-Gordon coefficient. Thus, the interbasis coefficients
in expansion (\ref{EXP2}) are given by
\begin{eqnarray}
\label{COEF3}
{\tilde W_{n m}^{n_1}(\nu)}
=
(-1)^{\frac{|m|-m+n_1}{2}}
\,
C_{\frac{n+\nu}{2}, \, \frac{\nu+m}{2}; \,
\frac{n+\nu}{2}, \, \frac{\nu-m}{2}}^{n_1+\nu, \, \nu}
\end{eqnarray}
and may be expressed in terms of the $_3F_2$ function through
(\ref{COEF1}).

Using the same method we could calculate the coefficients of
the interbasis expansion for the wave functions (\ref{WF3}) and
(\ref{WF1}). We have
\begin{eqnarray}
\label{EXP3}
\Psi_{l_1 l_2}(\theta'', \varphi'') =
\sum_{m = 0}^{n}
(-1)^{n+\frac{m}{2}}
W_{l_1 l_2}^{m}(\nu)
\Psi_{n m}(\theta, \varphi)
\end{eqnarray}
where the coefficients $W_{l_1 l_2}^{m}(\nu)$ are given by formulae
(\ref{COEF1}) or (\ref{COEF2}) by replacing the quantum number
$n_i\rightarrow l_i$.

The last interbasis expansion between two spherical wave functions
(\ref{WF3}) and (\ref{WF2}) can be constructed by using
equations (\ref{EXP3}) and (\ref{EXP2})
\begin{eqnarray}
\label{EXP4}
\Psi_{l_1 l_2}(\theta'', \varphi'')
=
\sum_{n_1 = 0}^{n}
U_{l_1 l_2}^{n_1}(\nu)
\Psi_{n_1 n_2}(\theta', \varphi'),
\end{eqnarray}
where
\begin{eqnarray}
U_{l_1 l_2}^{n_1}(\nu)=
(-1)^{l_2+\frac{l_1+n_1}{2}}
\sum_{m = - n}^{n}
(-1)^{\frac{m}{2}}
C_{\frac{n+\nu}{2}, \, \frac{\nu+m}{2}; \,
\frac{n+\nu}{2}, \, \frac{\nu-m}{2}}^{l_1+\nu, \, \nu}
C_{\frac{n+\nu}{2}, \, \frac{\nu+m}{2}; \,
\frac{n+\nu}{2}, \, \frac{\nu-m}{2}}^{n_1+\nu, \, \nu}
\end{eqnarray}
Finally, note that direct methods of calculation of the
coefficients in expansion (\ref{EXP4}) give us the hypergeometrical
function $_4F_3$ from the unit argument.
end of last section

\section*{Acknowledgment(s)}
The authors thank V.M.Ter-Antonyan and L.G.Mardoyan for
interesting discussions. One of the authors (G.~P.) thankful
to the Organizers of the II International Workshop
"Lie Theory and its Application in Physics"
for financial support and very kind hospitality.

\end{document}